\documentstyle{mn}
\def\ebv{E(B$-$V)}

\def\vi{V$-$I}
\def\bv{B$-$V}
\def\gsim{\;\lower.6ex\hbox{$\sim$}\kern-7.75pt\raise.65ex\hbox{$>$}\;}
\def\lsim{\;\lower.6ex\hbox{$\sim$}\kern-7.75pt\raise.65ex\hbox{$<$}\;}

\title[Collinder 261]{CCD photometry of the old open cluster 
 Collinder 261\thanks{Based on  observations made at ESO telescopes, La Silla, Chile}}

\author[Gozzoli et al.]{E. Gozzoli$^1$, M. Tosi$^2$, G. Marconi$^3$, A. Bragaglia$^2$\\
 $^1$ Dipartimento di Astronomia, Universit\`a di Bologna, via Zamboni 33,
 I-40126 Bologna, Italy, 
      e-mail gozzoli@astbo3.bo.astro.it \\
 $^2$ Osservatorio Astronomico, via Zamboni 33,
 I-40126 Bologna, Italy, 
      e-mail tosi@astbo3.bo.astro.it, angela@astbo3.bo.astro.it\\
 $^3$ Osservatorio Astronomico, Via Osservatorio 5,
      I-00040 Monte Porzio (Roma), Italy, 
      e-mail marconi@coma.mporzio.astro.it}
\date{}

\begin{document}
\maketitle

\begin{abstract}
 We present UBVI photometry for the old open cluster Collinder 261. From
 comparison of the observed colour-magnitude diagrams with simulations
 based on stellar evolutionary models we derive in a self consistent
 way reddening, distance, and age of the cluster:  \ebv=0.25$-$0.34, 
(m-M)$_0$=11.7-12.0, and age=7$-$8 Gyr or 9$-$11 Gyr, 
depending on the adopted stellar tracks. The models in better agreement with
the data turn out to have metallicity at most solar.
\end{abstract}


\begin{figure*}
\vspace{14cm}
\caption{Map of the observed regions of Cr 261. The points are scaled with
the star magnitude and represent the output of our photometry, so some
(bright) stars could be missing. The area is a mosaic of 5 CCDs and the
circle has a radius of 3.5 arcmin. Reference stars are indicated with numbers,
and their coordinates are given in Table 3.} \label{fig-1}
\end{figure*}

\section{Introduction}

Galactic open clusters are excellent tools to study the evolution of our Galaxy 
from the chemical and structural points of view. They provide information 
on the chemical abundances in the disk and relative radial gradients (e.g. 
Janes 1979, Panagia and Tosi 1981, Friel and Janes 1993), 
on the interactions between thin and thick disks (e.g. Sandage 1988), on 
the average radial velocities and stellar ages at different galactic 
locations (e.g. Janes and Phelps 1994), on the absolute age of the disk. 
This latter could also be obtained from isolated field stars, e.g.
studying the White Dwarfs 
luminosity function at its cool end; the actual value is still 
uncertain, varying from about 6.5 to 13 Gyr due to different models for the 
White Dwarfs cores and to different treatments of the cooling 
and crystallization processes (see for example Winget et al. 1987, Isern et 
al. 1995, Wood 1995), but the oldest ages (more than about 9 Gyr) seem 
to be preferred.
This would imply some kind of temporal continuity between the formation of disk 
and halo, since the youngest halo clusters are only slightly older than
this age (see e.g. Buonanno et al. 1994, or Chaboyer at al. 1996).
Besides this method, we are able to reliably date only star clusters,
and open clusters represent the only class of objects covering
both a large range of distances (several kpc around the sun) and a large range
of ages (from a few Myr up to $\sim$10 Gyr) and can therefore provide key
constraints to galactic evolution theories (e.g. Tosi 1995). 
To this aim, it is however crucial
that the observational data be very accurate and homogeneously treated to
avoid misleading effects (see also Carraro and Chiosi 1994).

In order to study in detail the metallicity and age distribution of open
clusters with galactocentric distance, we have undertaken a project to 
analyse with the required accuracy a sample of open clusters 
located at different galactic radii and supposed to have different ages and 
metallicities. Deep CCD photometry is taken and properly analysed for each 
of the examined clusters. Age, metallicity, reddening and distance modulus are 
derived from the resulting colour-magnitude diagrams (CMDs) and luminosity 
functions (LFs) through the comparison with the corresponding CMDs and LFs
generated by a numerical code for MonteCarlo simulations based on stellar 
evolution tracks and taking into account theoretical and observational 
uncertainties (Tosi et al. 1991).  
These simulations have proved to be much more powerful than
the classical isochrone fitting method to study the evolutionary
status of the analysed region both in galactic open clusters (Bonifazi et al.
1990) and in nearby irregular galaxies (Marconi et al. 1995). 
As an interesting by-product of our method we can evaluate the effects
connected to the adoption of different stellar
evolution models. 

So far we have presented the results on the young metal rich cluster NGC 7790
(Romeo et al. 1989) and the old metal poor cluster NGC 2243 (Bonifazi et al. 
1990) and will shortly present results on the old metal poor clusters NGC 2506
and NGC 6253 and the young cluster NGC 6603.

The galactic cluster Collinder 261 (Cr 261, C1234$-$682: 
$\alpha_{1950} = 12^h 34^m 54^s, \delta_{1950} = -68^o 12'; l=301^o.66,
b=-5^o.64$)  has been found 
old by Phelps et al. (1994) who find it to be at least as old as NGC 6791. 
Friel et al. (1995) consider it among the oldest open clusters and derive 
from moderate resolution spectroscopy a metallicity [Fe/H]=$-$0.14. On the other
hand, Cr 261 has been found old but metal rich by the recent studies of
Kaluzny et al. (1995) and Mazur et al. (1995, hereinafter MKK).
Here we present deep CCD photometry of the
cluster in the UBVI bands, from which we derive our own estimates of age,
metallicity, distance and reddening.

In Section 2 we present the observations and data reductions, in Section 3
we introduce the obtained CMDs, in Section 4
we address the cluster parameters obtained by simulations based on three
different classes of stellar models. The results are discussed in Section 5
in the context of structure and evolution of the galactic disk.
\begin{table*}
\caption{Journal of observations for the 5 fields centered on Cr 261.
 Equatorial coordinates of the centres of the fields are referred
 to 2000.0 and all exposure times are in seconds.}
\begin{tabular}{cllllll}
\hline
 \multicolumn{1}{c}{Field}  &\multicolumn{1}{c}{Coordinates}
&\multicolumn{1}{c}{Date (UT)}
&\multicolumn{1}{c}{U} &\multicolumn{1}{c}{B} 
&\multicolumn{1}{c}{V} &\multicolumn{1}{c}{I} \\
\hline
C & 12 37 58 $-$68 22 56 & March 4 & 1800 & 1200,60 & 600,60 & 60,60,600\\
  &                    & March 8 &      & 1200,900 & 720,720 & 720,720 \\
N & 12 38 20 $-$68 17 14 & March 6 &      & 1200,60  &   &30,60,60,720\\
  &                    & March 7 &      & 1500,10,60 & 60,60,10,840 & 10\\
E & 12 39 08 $-$68 22 12 & March 7 &      & 1200,1200,60 & 720,60,10 & 10,10,60,600,10 \\
S & 12 37 58 $-$68 28 24 & March 4 &      & 120,600  &600 & 60,600 \\
  &                    & March 10&      & 1500,60,1200 &20,720,20,600 & 720,20,900\\
W & 12 37 41 $-$68 22 12 & March 5 &      & 1200,60 &60,600 & 600,40 \\            
\hline
\end{tabular}

\begin{tabular}{ll}
\multicolumn{2}{c}{Standard areas observed }\\
\hline
Rubin 149   & March 4, 5, 8\\
PG 0918+029 & March 4, 6, 7, 8\\
PG 1047+003 & March 5, 7, 10\\
PG 1323$-$086 & March 6, 8, 10\\
PG 1633+099 & March 5, 7\\
\hline
\end{tabular}

\begin{tabular}{lc lc lc} 
\multicolumn{6}{c}{Seeing excursion for each night, in arcsec}\\
\hline
Date    & seeing  &Date    & seeing   &Date      & seeing \\
March 4 & 1.0-1.6 &March 5 & 1.2-1.4  & March 6  & 1.0-1.2 \\
March 7 & 1.2-1.7 &March 8 & 1.0-1.3  &March 10  & 0.9-1.1 \\
\hline
\end{tabular}
\end{table*}

\section{Observations and data reduction}

The cluster was observed with the direct camera of the Danish 1.54m telescope 
(La Silla), mounting a Tek 1024$\times$1024 CCD (\#28, scale 0.377 arcsec/pix).
All observations were done on March 4-10, 1995 (UT); at least
one night was of excellent photometric quality, while cirri were 
occasionally present during the others. 
Seeing conditions were quite good for the site/telescope and the seeing 
excursion for each night is given in Table 1.

We covered the cluster
with 5 partially overlapping fields (see Figure~\ref{fig-1}), one right on the
cluster centre, and four others North, East, South and West of it,
extending to about 8 arcmin from the centre. The North frame was originally
intended for field stars decontamination, since its upper part lies beyond
the radius assigned to this cluster in literature.
During an ensuing observing run, we also took a few frames with the Dutch
telescope (field of view 3$\times$3 arcmin), 
2 positioned to the North and 2 to the
South of our field, plus another one with the Danish, to the East, which we
did not analyse with the others because of the much worse seeing conditions,
but only used to define a control external field (see Section 3), reaching
out to about 13 arcmin from the cluster centre.
The centre was observed in the UBVI Johnson - Cousins filters, while the other
fields were only observed in BVI. We have both short and long exposures,
so we cover with the required precision 
from the red giants region to about 4 magnitudes fainter 
than the main sequence turn-off (TO).
Table 1 gives a journal of observations: column 1 indicates the field,
column 2 the field centre coordinates, column 3 the date (UT), while columns 
4-7 give the exposure times for each field/filter combination.

Several standard stars fields 
chosen from Landolt (1992) 
were taken every night; they are also indicated in Table 1.
Table 2 gives pixel and equatorial coordinates for the 7 reference stars
indicated in Figure~\ref{fig-1}.

All reductions were done using IRAF\footnote{IRAF is distributed by the NOAO, 
which are operated by AURA, under contract with NSF}.
The frames were 
trimmed, bias subtracted and flat fielded, then analysed with 
DAOPHOT-II (Stetson 1987, 1992)
in the standard way. Stars were found in the deepest I frame for each field; 
all other images of the same field
were then aligned and the same list of stars was used to measure instrumental
magnitudes using PSF fitting (Allstar). We reduced each frame separately,
without adding deep exposures, then we averaged the measures.
The central field, taken on March 4, was used as reference
for calibrations and for joining all other fields, which was done using
all the stars in common.

Aperture correction was applied to the ``cluster'' stars to compensate for
loss of light due to PSF fitting: about 15 isolated stars per frame were
measured with aperture photometry and a correction 
between the aperture and PSF photometry was found, to be applied to all the 
other stars in the same frame. This correction varies from frame to frame,
but is usually of the order of about 0.2--0.3 mag, with a low dispersion 
around the mean value for each image.

Standard stars were measured using aperture photometry;
we derive an average calibration throughout all nights, given by the following
transformation equations:

\[ {\rm B} = b  +  0.182 (\pm 0.017) \cdot (b-v) - 2.573 (\pm 0.020) \]
\[ {\rm V} = v  +  0.032 (\pm 0.008) \cdot (b-v) - 1.803 (\pm 0.009) \]
\[ {\rm V} = v  +  0.028 (\pm 0.007) \cdot (v-i) - 1.762 (\pm 0.016) \]
\[ {\rm ~I} = i~  - 0.009 (\pm 0.007) \cdot (v-i) - 2.594 (\pm 0.017) \]

\begin{figure*}
\vspace{13.8cm}
\caption{Left panels: magnitude residuals for the standard stars with the
adopted calibrations. Right panels: distribution with magnitude of the
photometric error index $\sigma$} \label{fig-2}
\end{figure*}

Here $b,v,i$ are instrumental magnitudes,  and B,V,I the corresponding
Johnson and Cousins magnitudes. 
To correct for extinction we used the average, during the whole observing run,
of the extinction coefficients taken from the database maintained by J. Burki
(Geneva Obs.) on the ESO/La Silla archive, accessible through {\tt www} 
(http://arch-http.hq.eso.org).
The magnitude residuals resulting from these calibrations are shown in the left
panels of Figure~\ref{fig-2}, while the photometric error index $\sigma$ given
by DAOPHOT of all the objects detected in the various bands is plotted in the
right panels. 

We decided to use the standards observed in all the nights rather than only 
those of the best night, because,
as can be seen from Figure~\ref{fig-2}, this only results in a slight
increase of the scatter about the mean relation, without introducing any
systematic trend and allows, instead, a safer calibration of all the frames.
We then feel quite confident that our calibration is able
to transform into one uniform system data taken during the observing run.

In all cases when V was calibrated both from (B--V) and (V--I), the final
assumed V mag is the weighted average of the 2 values.
The overlapping regions were used to homogenize all measurements in order
to produce a single output of multiband magnitudes for about 19000 objects.
The magnitude differences for stars measured in more than one field turned out 
to be always around 0.03 mag in I and to vary between 0
and 0.1 in B and V (except for one single case where it reaches 0.2).
In all cases, the magnitudes of the central field, taken in better photometric 
conditions, have been assigned a larger weight when averaging the different
derived values.
Only 11243 objects were detected in all the three B, V and I bands with 
an acceptable error ($\sigma \leq$0.1 mag), and 2779 lie
within a radius of 3.5$^\prime$. A table with the B, V, I magnitudes and
pixel coordinates of these 2779 stars is available electronically from the 
authors.

It was not possible to define a satisfactory calibration for the U filter, so 
we adopted the following procedure to obtain ``almost-calibrated'' U magnitudes.
We selected stars clearly belonging to the cluster main sequence (MS), of known
(B--V) colour; adopting for the reddening an average of E(B--V) = 0.3 (see 
section 4), we have applied the relation between (B--V)$_0$ and (U--B)$_0$ for 
main sequence stars (Lang 1992, p. 149) to get the U magnitude for each of 
these objects. This small sample of stars was then
used to translate the instrumental $u$ magnitudes to Johnson U for all the 
objects in the central field. We are aware of the crudeness of this 
method and only show the CMDs involving these U mag for morphology sake.

To assess the completeness degree of our measurements we used the DAOPHOT task
Addstar to add artificial stars (randomly in position, but distributed in
magnitude and colours as the measured objects) to the deepest B, V and I frames.
We reduced them again in the same way as before and simply counting the
recovered objects we could estimate the completeness at each magnitude level.
This was done 3 times per filter and the results are presented 
in Table 3. 

\begin{table}
\begin{center}
\caption{Pixel and equatorial coordinates (equinox 2000.0) for 7 reference
 stars.}
\begin{tabular}{crrrr}
\hline
 \multicolumn{1}{c}{n} &\multicolumn{1}{c}{X} &\multicolumn{1}{c}{Y} 
&\multicolumn{1}{c}{$\alpha$} &\multicolumn{1}{c}{$\delta$} \\
\multicolumn{1}{c}{} &\multicolumn{2}{c}{(pixel)} &\multicolumn{2}{c}{(2000.0)} \\
\hline
1 &-841.0 & 478.9 & 12 37 48.9 & $-$68 31 15.0 \\
2 & 622.0 &-346.3 & 12 36 53.6 & $-$68 22 10.8 \\
3 & 145.1 &1494.0 & 12 38 57.2 & $-$68 25 10.9 \\
4 &1594.0 &1113.0 & 12 38 31.7 & $-$68 16 12.9 \\
5 &1604.0 & 203.1 & 12 37 31.1 & $-$68 16 07.1 \\
6 & -29.3 &-104.7 & 12 37 09.6 & $-$68 26 13.4 \\
7 & 877.2 &1933.0 & 12 39 26.9 & $-$68 20 39.8\\ 
\hline
\end{tabular}
\end{center}
\end{table}

\begin{table}
\caption{Completeness of our measurements at various magnitude levels;
 also indicated is the standard deviation from the mean of the three 
 experiments.
 For comparison, the TO is at about V=16.7, B$-$V=0.85, V$-$I=0.95}
\begin{center}
\begin{tabular}{crrrrrr}
\hline
 \multicolumn{1}{c}{mag} &\multicolumn{1}{c}{comp} &\multicolumn{1}{c}{$\sigma$}
&\multicolumn{1}{c}{comp} &\multicolumn{1}{c}{$\sigma$} 
&\multicolumn{1}{c}{comp} &\multicolumn{1}{c}{$\sigma$} \\
 \multicolumn{1}{c}{} &\multicolumn{1}{c}{B} &\multicolumn{1}{c}{B}
&\multicolumn{1}{c}{V} &\multicolumn{1}{c}{V} 
&\multicolumn{1}{c}{I} &\multicolumn{1}{c}{I} \\
\hline
    14  & 100.0\% &0.0\% & 100.0\% &0.0\%  &100.0\% &0.0\% \\   
    15  & 100.0\% &0.0\% & 100.0\% &0.0\%  & 95.7\% &7.5\% \\   
    16  & 100.0\% &0.0\% &  97.7\% &4.0\%  & 91.0\% &6.6\% \\   
    17  &  95.3\% &4.0\% &  93.3\% &1.2\%  & 88.7\% &2.1\% \\   
    18  &  88.0\% &7.0\% &  89.7\% &2.5\%  & 79.3\% &2.1\% \\   
    19  &  81.0\% &3.0\% &  85.3\% &2.1\%  & 64.3\% &5.7\% \\   
    20  &  72.0\% &6.9\% &  73.7\% &3.2\%  & 51.0\% &3.0\% \\   
    21  &  54.7\% &8.0\% &  60.3\% &9.0\%  & 29.3\% &3.5\% \\   
    22  &  42.0\% &6.1\% &  45.0\% &6.0\%  &        & \\   
    23  &   7.0\% &4.4\% &  21.7\% &5.0\%  &        & \\   
\hline
\end{tabular}
\end{center}
\end{table}

\section{The Colour Magnitude Diagrams}

Figures~\ref{fig-3} and ~\ref{fig-4} show the CMDs obtained from our 
reductions. As can be seen from
Figure~\ref{fig-3}, the cluster is well visible even when all field stars are
plotted. The main sequence TO is at V=16.7, \bv=0.85, \vi=0.95.
These values are in perfect agreement with those deduced from the only 
published calibrated CMDs; namely, Fig.4 of MKK who have BVI photometry, 
and Fig.24 of Phelphs et al. (1994), who only have VI. Our sequence is better
defined, being somewhat less dispersed, and is clearly distinguishable
down to about 4 magnitudes fainter than the TO.
A few red bright stars are visible in both CMDs (around V=14, and \bv=1.45
or \vi=1.4), and, as done by Phelps et al. (1994) and MKK, 
we assign them to the red giant clump, corresponding to the core-He burning
phase. The magnitude distance from the red clump to the TO is therefore
$\delta$V$\lsim$ 2.7. This large value and the
structure of the CMD indicate that Cr 261 is an old cluster. MKK show (their 
Fig.6) a few very bright and very red stars: we identified them in
our frames, but they were saturated even in the shortest exposures.
\begin{figure*}
\vspace{14cm}
\caption{CMDs of Cr 261. a) V $vs$ V$-$I: the three panels show the totality
of stars measured and the ones falling in a radius of 6 or 3.5 arcmin.
b) The same, but for V $vs$ B$-$V. The cluster is easily distinguishable
in all panels, as is the strong field contamination.}
\label{fig-3}
\end{figure*}

As already said, we have U measurements only for the central field.
The classical B$-$V $vs$ U$-$B plane could
not be used to determine the reddening, since the TO stars are
too cold, and anyway our U calibration is not of the
best quality. Figure~\ref{fig-4} presents 
the CMDs involving the U band; in all three CMDs the MS is well defined, and we 
can clearly see a lot of blue straggler stars. Furthermore,
the subgiant/red giant branch and the red clump are quite apparent.
Looking in particular at the U $vs$ (U--V) diagram, a hint of binary sequence
may be seen: it lies just about 0.7 mag brighter than the MS ridge line.
The other two CMDs are not so useful in this case.
To better assess the significance of this binary sequence 
we built histograms in colour at different magnitude levels:
unfortunately only in one interval a secondary peak due to the possible 
binaries can be clearly seen. This is not enough to rule out
the possibility of a binary sequence;
furthermore, MKK found a large number of
eclipsing binary systems in Cr 261, and several of them lie in the
right region of the CMD (see their Fig.11). 
\begin{figure*}
\vspace{8cm}
\caption{CMDs of Cr 261 for the central field. a) U $vs$ U$-$B; 
b) U $vs$ U$-$V (note the hint of binary sequence on the right of the MS); 
c) U $vs$ U$-$I}
\label{fig-4}
\end{figure*}

Deconvolution of the cluster from the field stars is quite difficult,
if not flatly impossible, since the latter dominate: they much outnumber 
the cluster stars even in the central part (radius about 3.5 arcmin).
Unfortunately, Cr 261 was given in the literature a much smaller radius (5 
arcmin diameter, Lang 1992) than it actually has, so even our 
nominal external field contains several cluster members. 
The same problem was faced by MKK who observed a region of 16.5 by 16.5
arcmin centered on Cr 261, but were not able to define a convincing 
external control field for decontamination, and concluded that the cluster
radius must be at least 6 arcmin.
We decided to build an artificial ``external field'',  using the 4 Dutch
and the easternmost Danish fields. Of them we retained only the parts
lying more than 12 arcmin from the cluster centre.
We then scaled the number of stars to be taken into
account for decontamination with the ratio of the areas:
since the region from which we extracted the field stars is smaller than the 
central field we simulated the required number of stars, 
varying magnitudes and colours of the real ones following gaussian random
distributions within the ranges of the empirical photometric errors.
The resulting cluster/field stars ratio in the central area of radius 
3.5$^\prime$ is about 0.25 which means that the actual cluster members of
that area are approximately 800.

Figure~\ref{fig-5} (a) shows the CMD of the stars detected in the field between 
12 and 13 arcmin from the centre and Figure~\ref{fig-5} (b) the semi-empirical 
diagram resulting from the above procedure to scale the observed stars to the 
area of the central region of 3.5$^\prime$.
Due to the uncertainty of such procedure, we do not attempt any decontamination
of the cluster field: as will be seen in the next session, we consider it 
safer to add field stars to the simulations rather than cleaning the cluster 
CMD from them.

\begin{figure}
\vspace{13cm}
\caption{CMD for the external field at a distance between 12 and 13 arcmin from
the centre of Cr 261: (a), empirical diagram; (b), semi-empirical diagram
corresponding to the area of the central region within 3.5$^\prime$ from the
centre (see text for details).}
\label{fig-5}
\end{figure}

\section{Cluster parameters}

In their detailed study of binary stars in Cr 261, MKK suggest 
for this cluster an age of $\tau=6-$8 Gyr, a distance modulus around
13 mag and a reddening \ebv=0.22, all based on the similarity between the
features of its CMD and those of other old open clusters, in particular 
of Be 39. In addition, the discovery of extremely red giants with 
\vi$>$3 has led these authors to include Cr 261 among the clusters with
solar or higher metallicity (Kaluzny et al. 1995). This suggestion is
however in contrast with the [Fe/H]=$-$0.14 inferred by Friel et al. (1995) from
spectroscopic indices. 
Here, we follow a different procedure to estimate the values of these 
parameters from our own photometry.

We have applied to Cr 261 the method described by Tosi et al.
(1991) for nearby irregular galaxies, which is an expansion of the classical 
isochrone fitting and allows to
simultaneously derive the age, reddening and distance modulus of the cluster.
The method compares the observed CMDs of the system with synthetic CMDs
resulting from Monte Carlo simulations using the same
number of stars above the same limiting magnitude, and with the same
photometric errors and incompleteness factors in each magnitude bin as
derived from the photometric data. The theoretical diagrams are transformed
into the empirical (V,\bv) plane by finding the values of reddening and distance
modulus providing     the best agreement with the stellar distribution in the
observed CMD. 
If a synthetic CMD coincides with the observed one,
it thus yields the best age, reddening and distance modulus pertaining to 
the cluster in the framework of the adopted stellar evolution models. 
The resulting set of values, however, is not unique, 
because different stellar evolution models may produce different solutions.
For this reason, we have derived the cluster parameters with three different
data bases of stellar models, already known to predict rather different ages: 
this approach allows us to evaluate both the best parameter values  and the 
corresponding theoretical uncertainties.

The synthetic diagrams examined for Cr 261 are based on homogeneous
sets of stellar evolution tracks computed for several initial metallicities
and already proven by their authors to reproduce many observational constraints.
Our simulations have been performed with: a) the tracks with classical 
mixing length treatment of the convective zones computed by Castellani
et al. (1993, hereinafter FRANEC), b) the tracks with overshooting from 
convective cores by the Padova group (hereinafter BBC), 
c) the tracks by D'Antona et al. (1992, hereinafter CM),
with the new convection treatment proposed by Canuto \& Mazzitelli (1991).
Within the framework of each group of stellar models, we have performed several
simulations for any reasonable combination of age, reddening and distance 
modulus, all of which have been compared with the empirical CMD and luminosity
functions of Cr 261. We describe below only the best cases, selected on 
the basis of these comparisons. The best cases for each of the three classes
of stellar models are shown in Figure~\ref{fig-6}; other cases, useful for a better
understanding of the cluster conditions, are shown in Figure~\ref{fig-7}.

Due to the large back/foreground contamination affecting Cr261, and for 
an easier
comparison with MKK's diagrams, we have limited our analysis to the stars
located within 3.5$^\prime$ from the cluster center. This selection does not 
introduce any bias in the derived parameter values, since the CMD of different 
regions of Cr 261 are totally equivalent to each other (see Figure~\ref{fig-3}).
The contamination in this central area is less severe, but unfortunately it
still prevents to derive detailed information on some of the cluster 
characteristics, such as the colour and luminosity functions of the redder
stars and the relative number of stars in the various evolutionary phases.

Of the 11243 stars detected in B, V and I bands with $\sigma \leq$0.1 
mag, 2779 lie within a radius of 3.5$^\prime$. The CMD of this selected
sample is shown in panel (g) of Figure~\ref{fig-6}. Unfortunately the majority of 
the objects are probably field stars and (see discussion in section 3)
only about 800 can be safely considered as cluster members. The synthetic
diagrams discussed below therefore assume the cluster to be populated by 800 
stars. However, for further test, we have also performed 
several MonteCarlo simulations based on higher membership and found
that an adopted population of 1000 stars, or more, definitely provides
CMDs with excessively crowded main sequence and TO regions,
when compared with the observational diagram. This means that our 
estimate of 800 cluster members situated in the inner 3.5$^\prime$, based
on a statistical subtraction of the external field, is consistent with the
number of stars appearing to populate both the MS and post-MS regions of the 
CMD.

We have not included binary stars in our simulations, because the field
contamination is too large to allow to derive any significant information on 
their cluster distribution from the observed CMDs of Cr 261.
We are however inclined to believe that the fraction of binaries is fairly 
large in this system, first because of the large number of blue stragglers,
second because of the spread in the MS and post-MS stellar distribution which
is much larger than expected from the photometric error, and third because
of the relatively large number, 45, of binary stars actually discovered by
MKK.

\begin{figure*}
\vspace{15.5cm}
\caption{CMDs for the inner 3.5$^\prime$ of Cr 261. The observational diagram
of the objects with $\sigma \leq0.1$ is in panel (g). The other panels
show the synthetic CMDs in better agreement with the data, with (bottom)
or without (top) the addition of the objects detected in the 
semi-empirical external field. Panels (a) and (d) refer to FRANEC models with 
$\tau$=7 Gyr, (m-M)$_0$=11.9 and \ebv=0.30; panels (b) and (e) to BBC models 
with $\tau$=11 Gyr, (m-M)$_0$=11.7 and \ebv=0.25; panels (c) and (f) to CM 
models with $\tau$=6 Gyr, (m-M)$_0$=12 and \ebv=0.34. All these models assume
solar metallicity.}
\label{fig-6}
\end{figure*}

\subsection{Results with FRANEC stellar models}

FRANEC sets of models follow the evolution of stars between 0.6 and 1 
M$_{\odot}$ 
in the central hydrogen burning phases and the evolution of stars between
1 and 9 M$_{\odot}$ up to the onset of thermal pulses on the asymptotic giant
branch. They have been computed for several values of initial helium and 
metal abundances: for the case of Cr 261, we have used the tracks with
Y and Z equal to (0.27, 0.01), (0.27, 0.02), (0.30, 0.02).\footnote{Note 
that these tracks have been computed with the LAOL Los Alamos opacities. 
According to the authors, the effect of using instead the most recent OPAL 
Livermore opacities corresponds only to assuming a slightly larger 
metallicity (Cassisi et al. 1993). Therefore, the metallicity of the FRANEC 
models mentioned above should actually be taken as about half of their 
nominal Z value.}
However, all the synthetic diagrams in better agreement with the data are found
to be based on the set with composition (0.27, 0.02).
Models with higher helium content in fact present a fairly brighter clump
during the core helium burning phase, which implies a larger magnitude
difference between clump and TO stars, inconsistent with that derived from
the observational data. Models with metallicity lower than solar 
present a rounder shape of the TO region and a more extended subgiant branch
(see the best case in Figure~\ref{fig-7} (c)), which make more difficult a 
good reproduction of the observed diagram.

\begin{figure*}
\vspace{10cm}
\caption{Synthetic CMDs for the central region of Cr 261. Clockwise from the
top-left panel: 
(a), FRANEC (0.27, 0.02), $\tau$=6 Gyr, (m-M)$_0$=12.0, \ebv=0.30; 
(b), FRANEC (0.27, 0.02), $\tau$=10 Gyr, (m-M)$_0$=11.7, \ebv=0.27; 
(c), FRANEC (0.27, 0.01), $\tau$=8 Gyr, (m-M)$_0$=11.7, \ebv=0.36; 
(d), BBC (0.28, 0.02), $\tau$=8 Gyr, (m-M)$_0$=11.8, \ebv=0.30; 
(e), BBC (0.35, 0.05), $\tau$=9 Gyr, (m-M)$_0$=11.7, \ebv=0.19; 
(f), BBC (0.28, 0.008), $\tau$=9 Gyr, (m-M)$_0$=11.7, \ebv=0.34.
}
\label{fig-7}
\end{figure*}

Figure~\ref{fig-6} (a) shows the CMD assuming $\tau$=7 Gyr, (m-M)$_0$=11.9 and
\ebv=0.30. The cluster luminosity, colour and stellar distribution in the
MS and TO regions are well reproduced by this model, as can be more easily
appreciated in Figure~\ref{fig-6} (d) where we have superimposed to the synthetic 
diagram the semi-empirical CMD of the external field, discussed in section 3.
Stars in the post-MS phases appear more concentrated in the synthetic CMD
than in the observational one. Nonetheless, their distribution is consistent 
with the data if one considers that Cr 261 is probably populated by 
a significant fraction of binary stars whose interactions during the giant
evolutionary phases may well be the cause of the observed spread in the 
CMD.

The same good result is obtained assuming $\tau$=8 Gyr, (m-M)$_0$=11.8 and
\ebv=0.30, and the two synthetic CMDs are in practice indistinguishable. 
Notice however that, since age, distance modulus and reddening all affect the 
apparent magnitude of the stars, to reproduce the observational diagram of 
Cr261 with an age higher than in the previous case, we must
assume a slightly smaller distance modulus to compensate the correspondingly 
lower intrinsic luminosity of the TO.

Models with $\tau >$8 Gyr are less satisfactory, because to keep on
fitting the MS and TO regions one needs to assign to the cluster reddenings
or distance moduli which make the synthetic red giant branch (RGB) and clump
definitely bluer than observed,  as shown in Figure~\ref{fig-7} (b), where the CMD
corresponding to 10 Gyr, \ebv=0.27 and (m-M)$_0$=11.7 is shown. In other words, 
for ages older than 8 Gyr, we do not find a
combination of \ebv ~and (m-M)$_0$ able to well reproduce the CMD features.

An age slightly younger than 7 Gyr might also be attributable to Cr 261, with
the same reddening and slightly larger distance. For instance, the model with 
$\tau$=6 Gyr, (m-M)$_0$=12.0 and \ebv=0.30, shown in 
Figure~\ref{fig-7} (a) still
reproduces most of the observational features of Cr261. However, at $\tau$=6 
Gyr the gap corresponding to the MS overall contraction phase, typical of
clusters some Gyr old (e.g. NGC 2243, see Bonifazi et al. 1990), 
is present in 
the synthetic CMD. Since the gap does not appear in the observational diagram, 
Cr 261 must be older than 6 Gyr. 
Notice that, had we used the classical isochrone fitting method to infer the
cluster parameters, we would not have noticed the gap and
would not have been able to put this lower limit to the age of
Cr 261. For even 
younger ages the shapes of both the MS and the TO regions change,
with the MS becoming more straight and the TO showing the hook 
typical of young systems. These modifications make such younger diagrams 
inconsistent with the data. 

In summary, adopting FRANEC evolutionary tracks, we find that Cr 261 has
an age between 7 and 8 Gyr, a distance modulus in the range 11.8-11.9 and
a reddening 0.3, and is consistent with an initial chemical composition
nominally solar but actually lower (see note), and consistent with the value
[Fe/H]$\simeq-$0.14 derived by Friel et al. (1995) from low dispersion spectra.

\subsection{Results with BBC stellar models}

The stellar evolution tracks computed by the Padova group take into account
the possible overshooting of convective regions out of the edges defined by 
the classical mixing length theories. They have been computed for the whole mass
range (i.e. between 0.5-0.8 and 100-120 M$_{\odot}$) and several initial 
metallicities. They reach the tip of the asymptotic giant branch or the 
ignition of the core C-O burning phase, depending on the initial stellar mass.
For Cr 261, we have tested the sets of tracks with Y and Z equal to 
(0.28, 0.008) by Alongi et al. (1993), (0.28, 0.02) by Bressan et al. (1994) 
and (0.352, 0.05) by Fagotto et al. (1994), available at the 
Strasbourg Data center.

Figure~\ref{fig-6} (b) shows one of the BBC synthetic CMDs in better agreement 
with the data. It assumes solar chemical composition, $\tau$=11 Gyr, \ebv=0.25
and (m-M)$_0$=11.7 and reproduces pretty well both the MS and post-MS colours,
luminosities and stellar distributions, as can be better appreciated in
panel (e) where the synthetic CMD overlaps that of the semi-empirical
external field. Equivalent results are obtained with $\tau$=9 Gyr, 
\ebv=0.27 and (m-M)$_0$=11.7 and with $\tau$=10 Gyr, \ebv=0.26 and 
(m-M)$_0$=11.7. However, 9 Gyr is the minimum age attributable to Cr 261
with this set of tracks, because at smaller $\tau$ they show a large MS
gap and a hooked TO, inconsistent with the observed shape, as already evident 
in the CMD in Figure~\ref{fig-7} (d), where the adopted parameters are
$\tau$=8 Gyr, \ebv=0.3 and (m-M)$_0$=11.8. 
On the other hand, the cluster might be as old as 12 Gyr, but the shape of
the MS and post-MS phases at this age starts to deviate from the 
observed ones to become more similar to those observed in globular clusters. For
instance, the subgiant branch runs at fairly constant luminosity and the base 
of the RGB is increasingly brighter than observed.

The same age (10$\pm$1 Gyr) and distance modulus (11.7) are obtained also 
using the more metal rich set of tracks. A significantly lower reddening
0.16$\leq$\ebv$\leq$0.19 is however required to compensate the redder 
intrinsic colours due to the higher metallicity. The problem with this set of 
tracks is that, probably because of the large helium content, the predicted 
red giant clump is more luminous than with the solar composition set, and 
0.5-1 magnitude brighter than observed in Cr 261 (see e.g. Figure~\ref{fig-7} (e) 
for $\tau$=9 Gyr, \ebv=0.19 and (m-M)$_0$=11.7)

An age of 9-10 Gyr and the same distance modulus 11.7 are inferred also 
with the lower metallicity set of models (Z=0.008). Since these tracks 
are intrinsically bluer, because of the lower metal content, they require
a slightly larger reddening \ebv=0.32-0.34 to fit the data. Figure~\ref{fig-7} (f)
shows the synthetic CMD with $\tau$=9 Gyr, \ebv=0.34 and (m-M)=11.7.
In these stellar tracks, the RGB spans a larger colour range than
in any other of the evolutionary sets considered here and this leads to
the shallower slope of the giant branch visible in Figure~\ref{fig-7} (f). Given 
the spread in the corresponding observational distribution, we are however
unable to evaluate what is the appropriate slope. Instead, we consider the 
comparison between this low metallicity tracks and the observational
diagram less satisfactory than that of the solar metallicity models because
of the shape of the TO region and the larger colour extension of the subgiant 
branch.

Therefore, we can conclude that with BBC models the age of Cr 261 is
$9\leq\tau\leq$11 Gyr and the distance modulus (m-M)$_0$=11.7, independently 
of the cluster metallicity. The solar composition set of tracks seems
the most appropriate and implies a reddening 0.25$-$0.27.

\subsection{Results with CM stellar models}

A few years ago, Canuto \& Mazzitelli (1991) have proposed a new approach to
the treatment of the stellar convective zones, alternative to the classic
mixing length theory, and D'Antona et al. (1992) have computed with such
approach the evolutionary tracks of stars
of initial mass between 0.65 and 1.5 M$_{\odot}$ and chemical abundance
Y=0.285 and Z=0.018, up to the core helium burning. These tracks show 
interesting differences in the relative
timescales, luminosities and temperatures of the various phases and we thus
consider it useful to compare their predictions with the observed CMD of Cr 261.

Figure~\ref{fig-6} (c) and (f) show the synthetic CMD based on the CM tracks in 
better agreement with our observational diagram. It assumes an age of 6 Gyr, 
(m-M)$_0$=12.0 and \ebv=0.34. Similar results are obtained with $\tau$=7 Gyr
if the lower TO luminosity is compensated by a lower reddening.
These tracks show a curvature of the MS just below the TO higher than with 
classic models and a slightly hotter luminosity minimum at the base of the
red giant branch. 
Unfortunately the field contamination is too large to 
allow any choice between the CM and classical cases and, within the 
uncertainties, both look in very good agreement with the data.

\section{Discussion}

Comparing observational data with the synthetic CMDs is useful not only to
measure fundamental quantities for the observed objects, but also to better
define the actual uncertainties associated with stellar evolutionary models. In
fact, despite the very different assumptions of the three sets of evolutionary
tracks adopted here, and the different solutions obtained from each of them, 
we are
unable to univocally and uncontroversially choose $the$ best fit among them:
considering the observational errors all three sets have at least one good
solution for Cr 261. This clearly translates into a ``theoretical uncertainty''
on the derived parameters, which affects most strongly the age.
Also, this implies that age estimates from a single set of isochrones
(beyond the fact that isochrone fitting is a less powerful tool than
comparison to synthetic CMDs) hide a much larger uncertainty than quoted,
since the latter only reflects the $internal$ error for the chosen set
of models. Finally, this also means than no ranking in age
can be truly believed if it has not been derived in a homogeneous way, as
already remarked by e.g. Carraro \& Chiosi (1994) and Friel (1995).

Taking into account the mentioned uncertainties, we prefer to give for Cr 261
not a simple ``best value'' for the quantities, but instead
the following ranges: 
7$\lsim\tau\lsim$11 Gyr, 0.25$\lsim$E(B-V)$\lsim$0.34, 
11.7$\lsim$(m-M)$_0\lsim$12.0.  Metallicity is about solar, with 
indication for being slightly lower.

How do our findings compare to the literature on Cr 261~? Phelps et al. (1994)
give a $\delta$V (difference in magnitude between red clump and TO) of 2.6
(in agreement with our $\lsim$2.7), second only to Be 17, and equal to that 
attributed to NGC 6791, for which the age estimates vary from about 5.5 to 10 
Gyr. An age ranking based on $\delta$V is not strictly monotonic and may be
rather unsafe in cases like Cr 261 where the red clump is poorly defined, but 
such large $\delta$V undoubtedly indicates very old ages. Phelps et al. are 
inclined towards the upper end of the age range and quote the 9 Gyr assigned 
to NGC 6791 by Garnavich et al. (1994) when discussing the age difference 
between the oldest open clusters and the younger globular clusters. They do 
not give any indication about metallicity or reddening.
Our age estimate is in good agreement with their suggestion: Cr 261 is 
surely among the oldest open clusters of our Galaxy.

MKK measure a $\delta$V of 2.4-2.5, comparable to that of NGC 188, implying a 
slightly younger age: their estimate is of about 6-8 Gyr. They find distance 
modulus and reddening, (m-M)$_0$=13.0, \ebv=0.22, from comparison to Be 39, but
admit that this is only a preliminary determination. Their estimate of 
metallicity (comparable to solar or higher) is based on the presence of very 
red giants, by analogy to NGC 6791. Our age estimate is consistent with MKK's,
but \ebv ~and distance modulus are not. We regard our estimate as more accurate
than theirs, since it was explicitly beyond the scope of their work to 
determine precise values for those quantities, and they only do it by 
comparison with other clusters and one single isochrone. Our values are 
systematically different from theirs, independently of the adopted set of 
evolutionary tracks. In other words, {\it none} of the synthetic CMDs is able
to reproduce the observational one assuming a distance modulus as large as 13.
The difference with MKK parameters is more significant if one considers that
a reddening as low as their 0.22 is acceptable in our simulations only if we 
adopt the BBC high metallicity tracks (morphologically inconsistent with the 
observed CMD) and (m-M)$_0$=11.7, i.e. much lower than their suggested 13.
Vice versa, our reddenning and distance modulus ranges are perfectly consistent
with the values, 0.33 and 12.04 respectively, derived by Janes \& Phelps (1994).

Concerning the chemical composition, we did not find any 
indication in favour of a metallicity higher than solar. 
On the contrary, our best fits have always been for solar metallicity;
furthermore, as noted before, for the FRANEC tracks the nominally-solar set
of models corresponds to a slightly lower effective metallicity. This is
in good agreement with the only metallicity spectroscopically determined 
([Fe/H]=--0.14, Friel et al. 1995).

Whatever its age, Cr 261 is clearly a disk object: its galactic coordinates, 
combined with the distance from the Sun derived from the distance modulus, 
place it just in close proximity of the Sagittarius spiral arm. What we are
seeing now is probably the surviving portion of a much bigger stellar
system which has had the time to spread around many but not all of its original
members, as suggested by the large spatial extension, but with very moderate
concentration, of the region containing stars with Cr 261 features.
Its very old 
age (greater than 6 Gyr, and possibly as high as 11 Gyr), close to that of
the youngest globulars, and the fact that it does not seem to be the oldest 
known open cluster, indicate that the disk contains an old population of 
relatively metal rich objects that must be accounted for by realistic galactic 
formation and evolution models, both dynamical and chemical.

\bigskip\bigskip
We wish to thank Laura Greggio for the code which has been the bulk of the 
numerical procedure for CMD simulations.
We are grateful to Franca D'Antona, Sandro Bressan and Oscar Straniero
for useful conversations on the evolutionary tracks.
This research has made use of the Simbad database, operated at CDS,
Strasbourg, France.

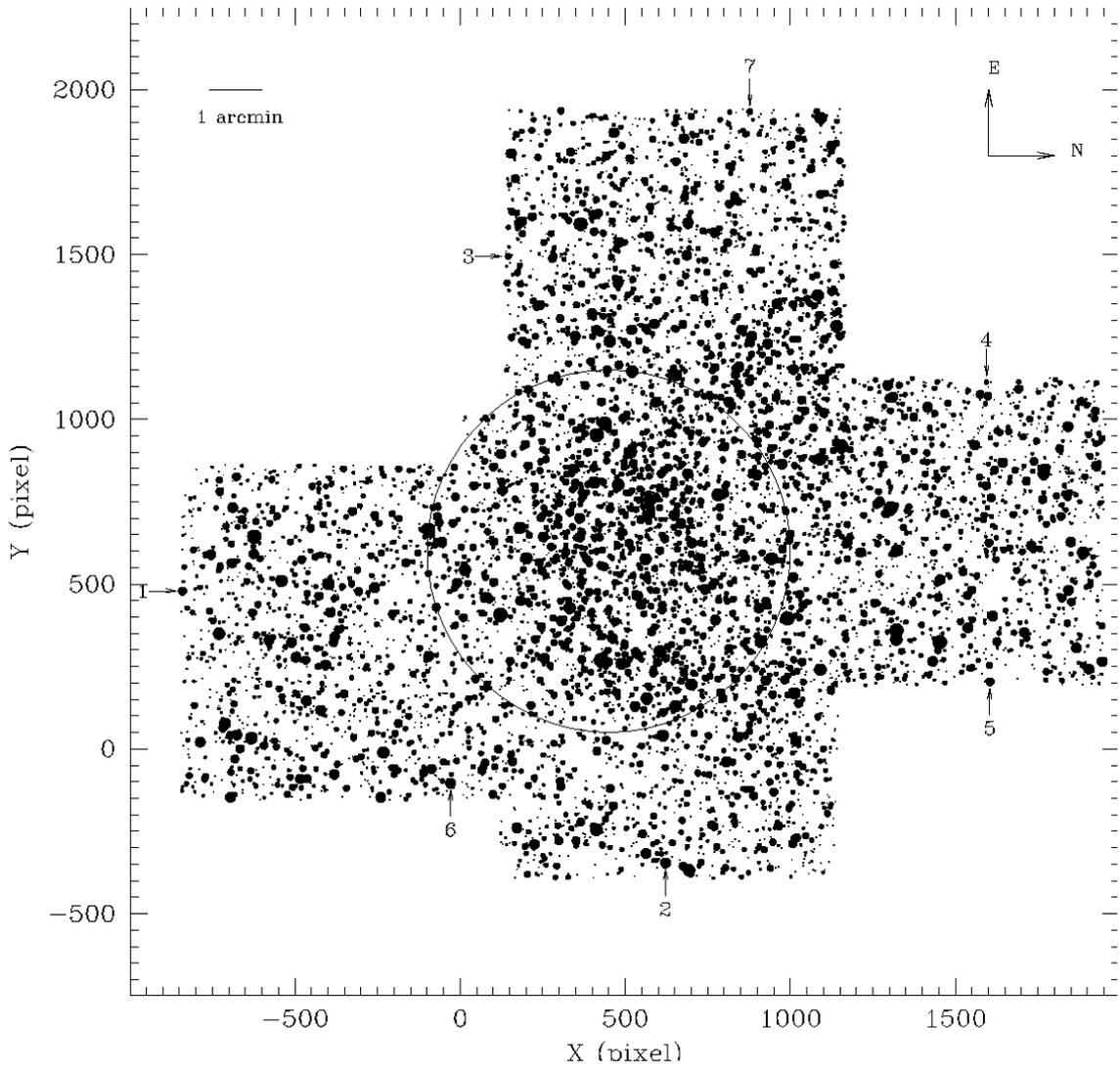
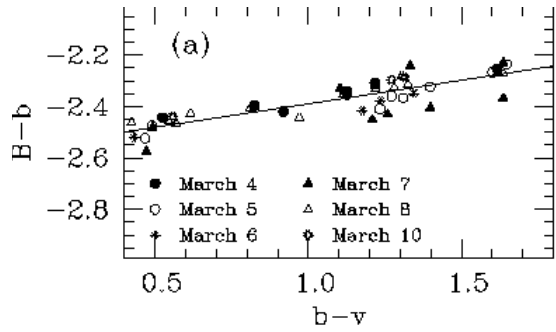
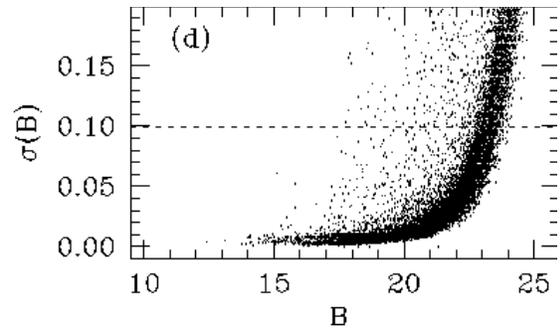
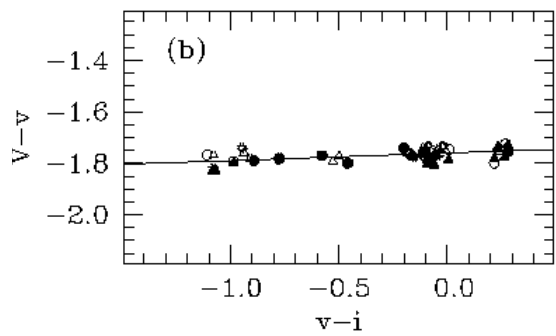
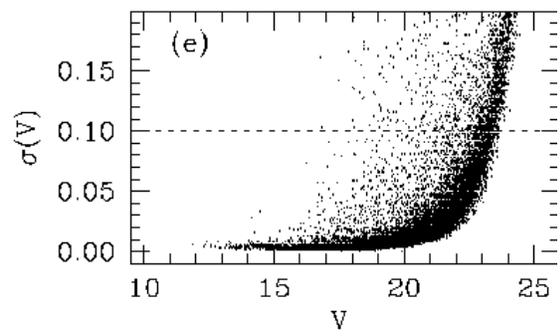
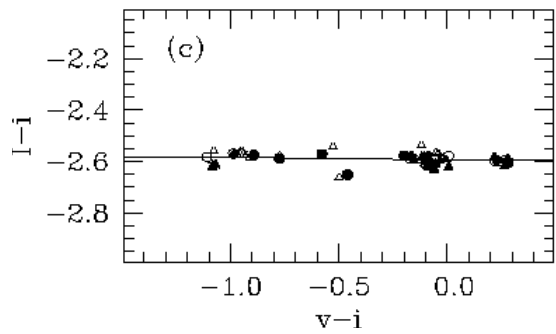
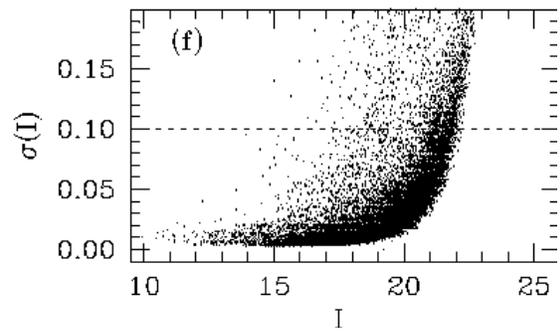
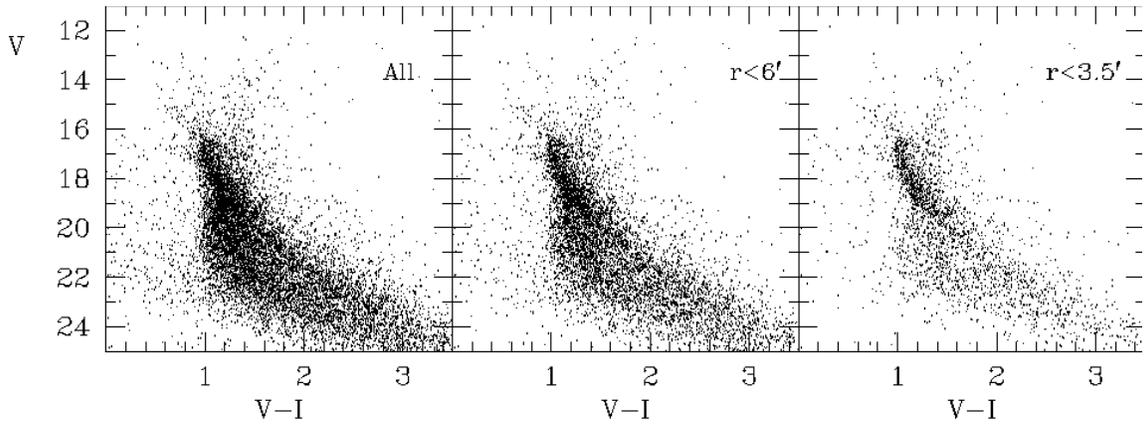
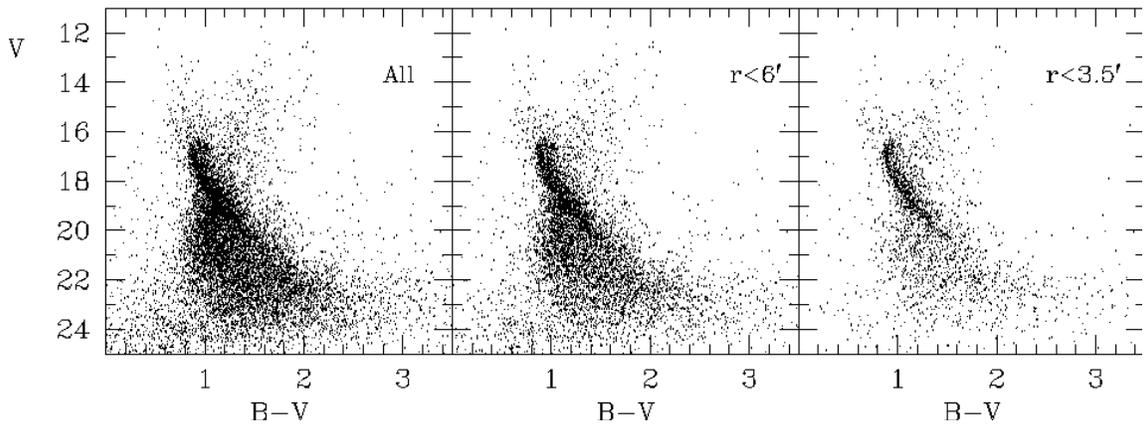
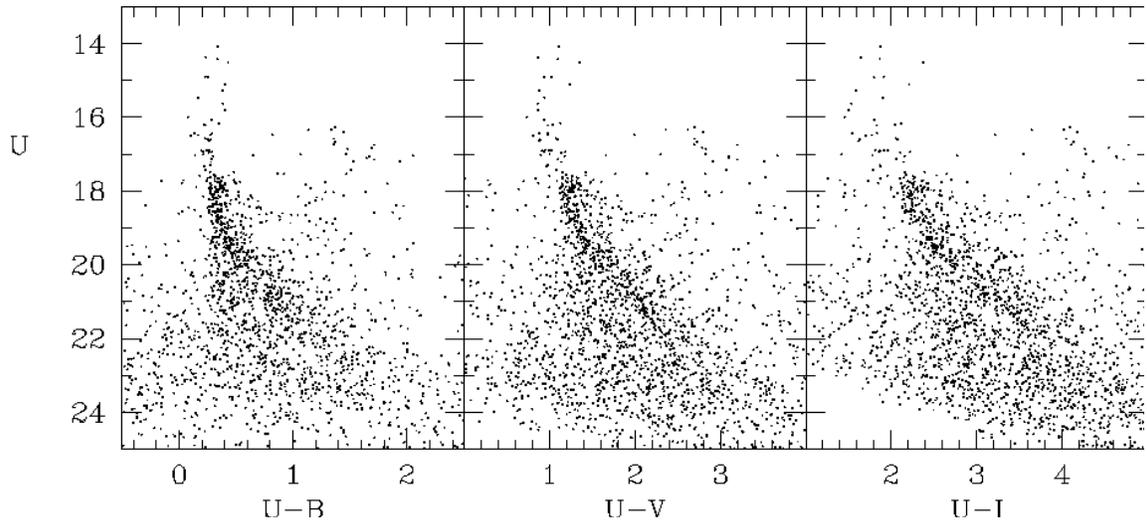
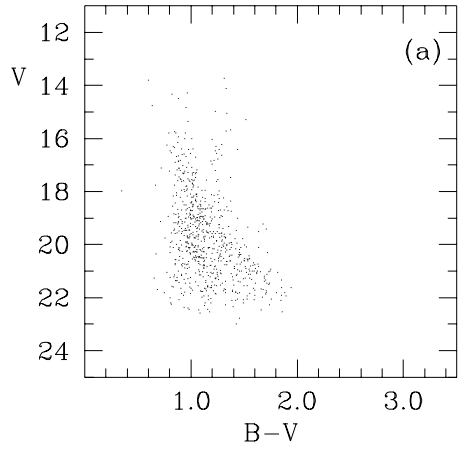
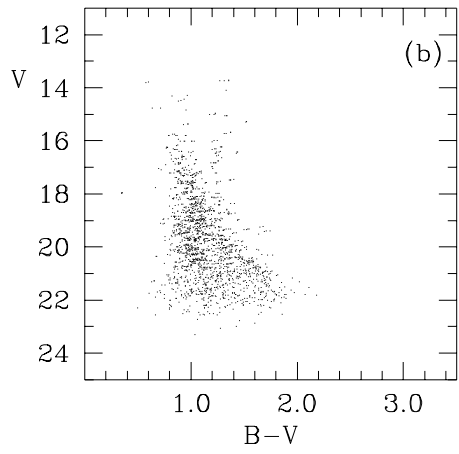
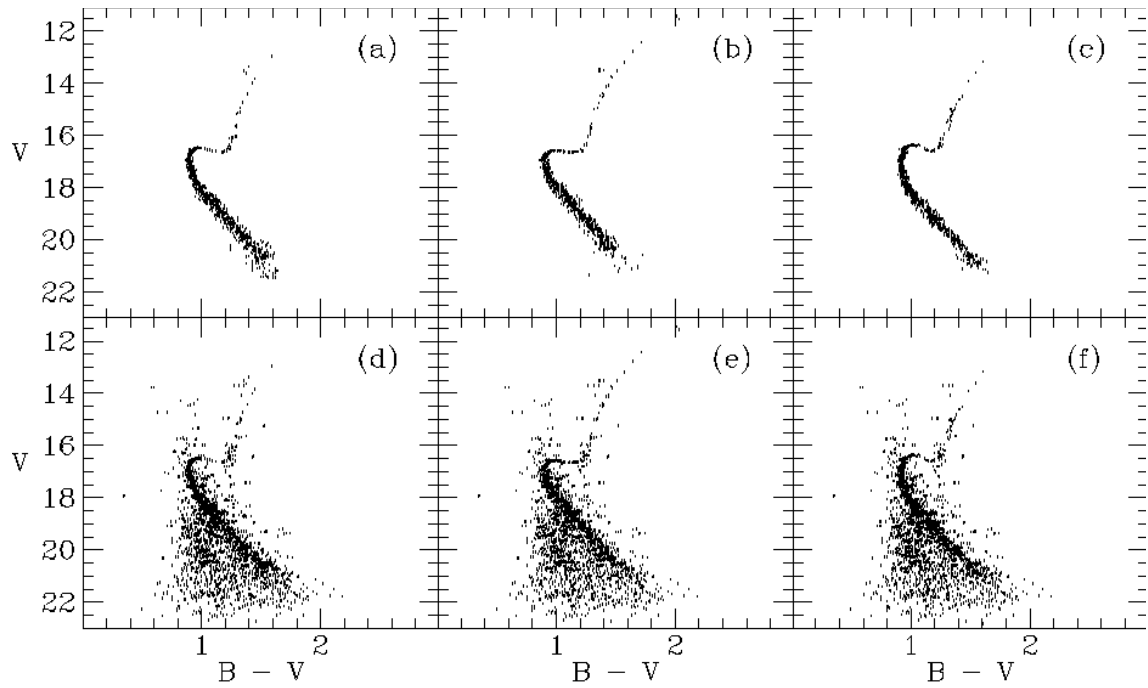
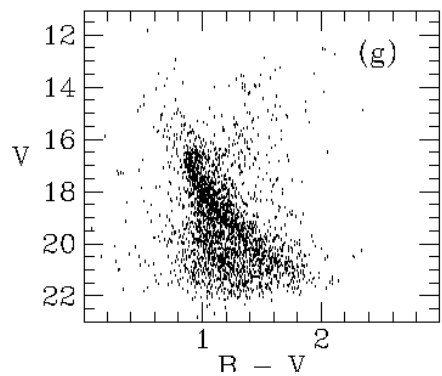
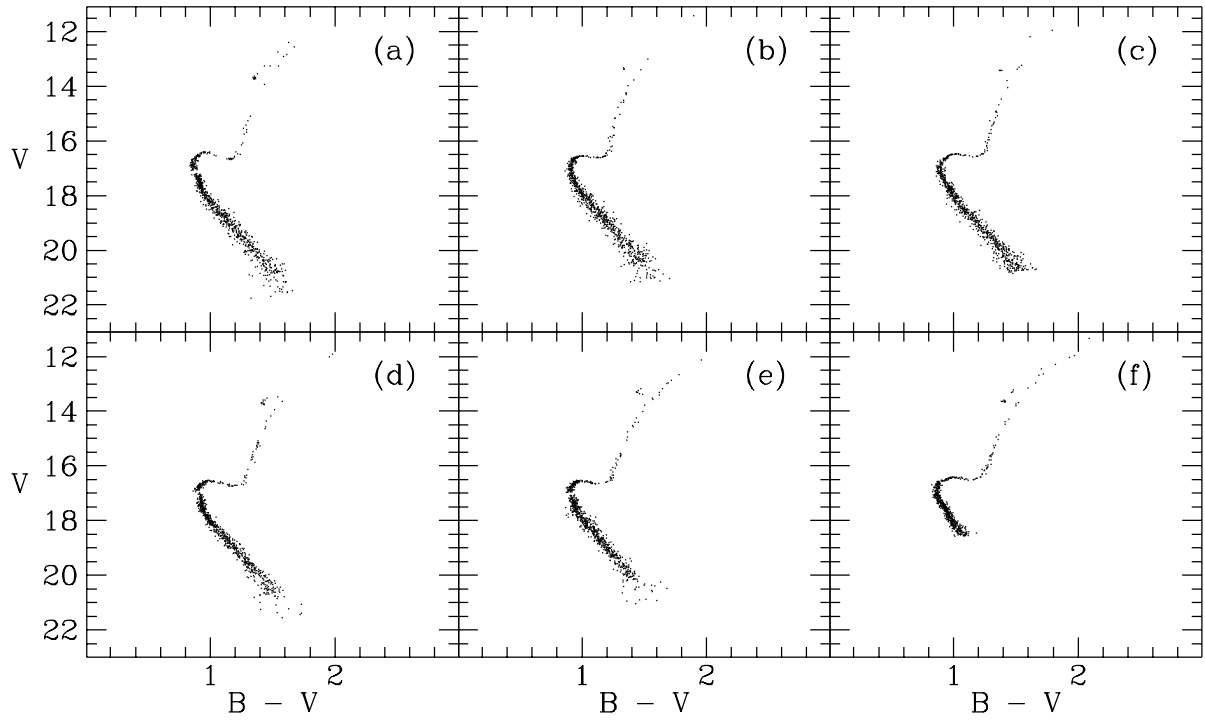
\end{document}